\documentclass[aip, sd, amsmath,amssymb, reprint]{revtex4-1}
\usepackage{graphicx}  
\usepackage{dcolumn}  
\usepackage{bm}  
\usepackage{dcolumn}  
\usepackage{amssymb}   
\usepackage{sidecap}
\usepackage[colorlinks=true, plainpages = false, citecolor = blue, urlcolor = blue, filecolor = blue]{hyperref}

\begin{document}

\preprint{AIP/123-QED}

\title[Applied Physics Letters]{Strained GaN Quantum-Well FETs on Single Crystal Bulk AlN Substrates}

\author{Meng Qi}
\author{Guowang Li}
\author{Satyaki Ganguly}
\author{Pei Zhao}
\author{Xiaodong Yan}
\author{Jai Verma}
\affiliation{Electrical Engineering, University of Notre Dame, Notre Dame, IN 46556}
\author{Bo Song}
\author{Mingda Zhu}
\author{Kazuki Nomoto}
\affiliation{Electrical and Computer Engineering, Cornell University, Ithaca, NY 14853}
\author{Huili (Grace) Xing}
\author{Debdeep Jena}
\email{djena@cornell.edu.}
\affiliation{Electrical Engineering, University of Notre Dame, Notre Dame, IN 46556}
\affiliation{Electrical and Computer Engineering, Cornell University, Ithaca, NY 14853}
\affiliation{Materials Science and Engineering, Cornell University, Ithaca, NY 14853}

\date{\today}

\begin{abstract}
We report the first realization of molecular beam epitaxy grown strained GaN quantum well field-effect transistors on single-crystal bulk AlN substrates.  The fabricated double heterostructure FETs exhibit a two-dimensional electron gas (2DEG) density in excess of 2$\times$10$^{13}$/cm$^{2}$.  Ohmic contacts to the 2DEG channel were formed by n$^{+}$ GaN MBE regrowth process, with a contact resistance of 0.13 $\Omega \cdot $mm.  Raman spectroscopy using the quantum well as an optical marker reveals the strain in the quantum well, and strain relaxation in the regrown GaN contacts.  A 65-nm-long rectangular-gate device showed a record high DC drain current drive of 2.0 A/mm and peak extrinsic transconductance of 250 mS/mm.  Small-signal RF performance of the device achieved current gain cutoff frequency $f_T \sim 120$ GHz.  The DC and RF performance demonstrate that bulk AlN substrates offer an attractive alternative platform for strained quantum well nitride transistors for future high-voltage and high-power microwave applications.  
\end{abstract}

\maketitle


State-of-art gallium nitride based electronic devices have demonstrated excellent performance in high-frequency and high-power applications \cite{iedm12_shinohara_hemt, edl12_yue_ganhemt_370GHz, edl12_palacios_ganhemt_300GHz, jjap13_yuanzheng_400GHz, drc12_denninghof_ganhemt_400GHz}.  These devices are on thick GaN buffer layers, most of which are on SiC substrates for efficient thermal dissipation.  The heteroepitaxially grown GaN layers inherently incorporate high density of dislocations (typically $\sim 10^9$/cm$^2$), which give rise to reliability issues and degrade breakdown characteristics.

In this letter we show that thin strained GaN quantum well double heterostructures on bulk AlN substrates offer an attractive alternative approach for high-performance nitride electronics.  To meet the scaling requirements for high-speed high-power RF applications, tight electrostatic control and quantum confinement of charge carriers are highly desired.   The wide direct band gap of AlN ($\sim$6.2 eV) and its large band offset with GaN offers the maximal vertical confinement of carriers in nitride channels.  The large polarization charge of AlN induces high density two-dimensional electron gases (2DEG) in the quantum well, which is desired for high current drive and lateral scaling of gate lengths.  The thermal conductivity of AlN, estimated \cite{jpcm87_slack_aln_thermal_conductivity_340WmK} to be as high as $\sim$340 W/m$\cdot$K can be  comparable\cite{shur_gan_properties} to that of SiC $\sim$370 W/m$\cdot$K, and offers the potential benefit of reducing thermal boundary resistance. Thus AlN simultaneously satisfies the conflicting requirements of high thermal conductivity and high electrical resistivity for high-power microwave electronics.  Importantly, the low dislocation density ($\sim 10^4$/cm$^2$) of single-crystal bulk AlN substrates has the potential for defect-free channels and barriers in principle, which is a promising prospective for improving thermal robustness, reliability, breakdown, and noise characteristics of these devices \cite{apl03_gan_hemt_on_aln_crystal_IS}.  The presence of a very thin GaN quantum well active region embedded in the AlN/GaN/AlN double heterostructure also enables selective optical-marker based Raman metrology of strain \cite{apl15_meng_raman} present in the active layers.

We recently reported the transport properties and device performance of AlN/GaN/AlN heterostructures, but all those studies were performed on AlN templates on sapphire \cite{apl14_guowang_gan_pHEMT, edl12_guowang_ganQW_nFET}.  In that work, a number of fundamentally new design paradigms enabled by these heterostructures were outlined.  In this letter we demonstrate the binary heterostructures on single crystal bulk AlN substrates.  The DC and RF performance of the FETs on bulk AlN are improved from the counterparts on sapphire.

\begin{figure}[t]
\includegraphics[scale=0.19]{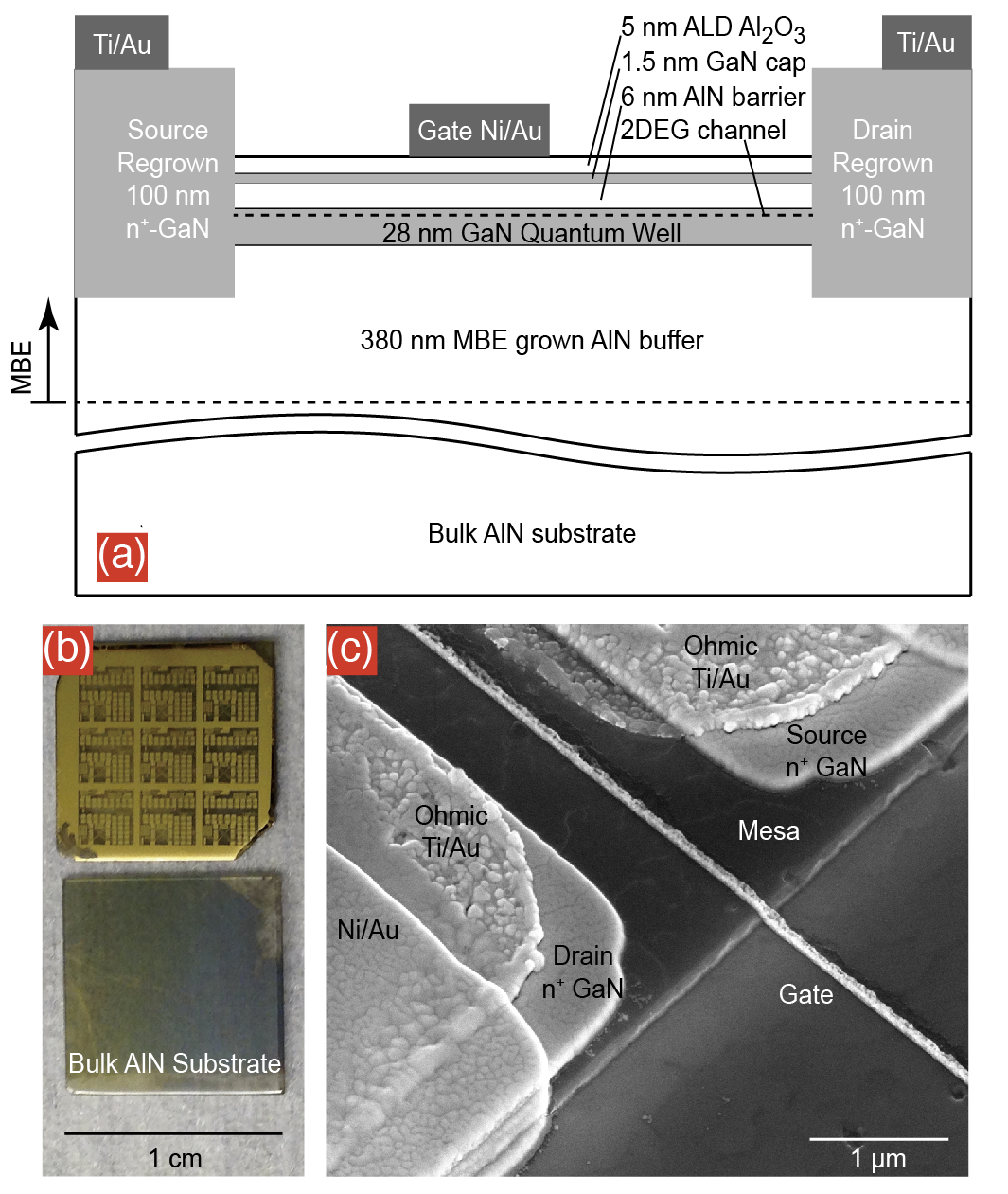}
\caption{\label{fig1} (a) Schematic cross-section layer structure of the AlN/GaN/GaN heterostructure FETs on bulk AlN substrates (not to scale).  (b)  Images of a grown and processed sample (top), and an unprocessed bulk AlN substrate (bottom).  (c)  SEM image of a finished short gate length GaN quantum well field-effect transistor.}
\end{figure}

A schematic cross-section of the device structure is shown in Fig. \ref{fig1} (a).  The AlN/GaN/AlN heterostructures were grown by RF plasma molecular beam epitaxy (MBE) on $\sim$400 $\mu$m-thick semi-insulating Aluminum-polar bulk AlN substrates.  Fig. \ref{fig1} (b) shows the bulk AlN crystal.  A 380-nm-thick unintentionally doped (UID) AlN buffer was epitaxially grown, followed by a 28 nm GaN quantum well where the 2DEG channel is located.  Then, a 6-nm-thick AlN layer was grown on the GaN quantum well as the top barrier, and capped with a 1.5-nm-thick GaN layer to protect the AlN surface from oxidation.  Hall effect measurements at room temperature showed a 2DEG density of $n_s \sim 2.8 \times 10^{13}$/cm$^2$, a room-temperature electron mobility of $\mu \sim 260$ cm$^2$/V$\cdot$s, and a corresponding sheet resistance of $ R_{sh} \sim 835$ $\Omega/\square$, with indications of defect generation during nucleation.  This sample was used for subsequent HEMT fabrication.  The mobility in such heterostructures can be significantly improved with modified nucleation conditions.  For example, in a separate 3 nm AlN/ 21 nm GaN/ AlN quantum well heterostructure grown on the same bulk AlN substrate with different nucleation conditions, a 2DEG mobility of $601/1380$ cm$^2$/V$\cdot$s, a sheet charge density of $3.2/2.6 \times 10^{13}$/cm$^2$, and a sheet resistance of $327/171$ $\Omega/\square$ at 300/77 K were observed.  These are the highest measured mobility and lowest sheet resistance for the AlN/GaN/AlN strained quantum well heterostructures on the AlN platform till date.  However the HEMT fabrication process on this lower sheet resistance sample was not successful.  At this stage, we do not completely understand the precise correlation between the nucleation conditions and the 2DEG transport properties.  Several mechanisms could be limiting the Hall-effect mobilities in the strained QW channels, such as rough heterointerfaces, Stark-effect scattering\cite{apl11_raj_stark_scattering}, and hole-drag due to a co-existence of a two-dimensional hole gas (2DHG) channel at the bottom GaN/AlN interface \cite{apl14_guowang_gan_pHEMT}.  A comprehensive study of transport is ongoing, and should help identify the lower measured electron mobilites in such heterostructures compared to 2DEGs on GaN substrates.

\begin{figure}[t]
\includegraphics[scale=0.4]{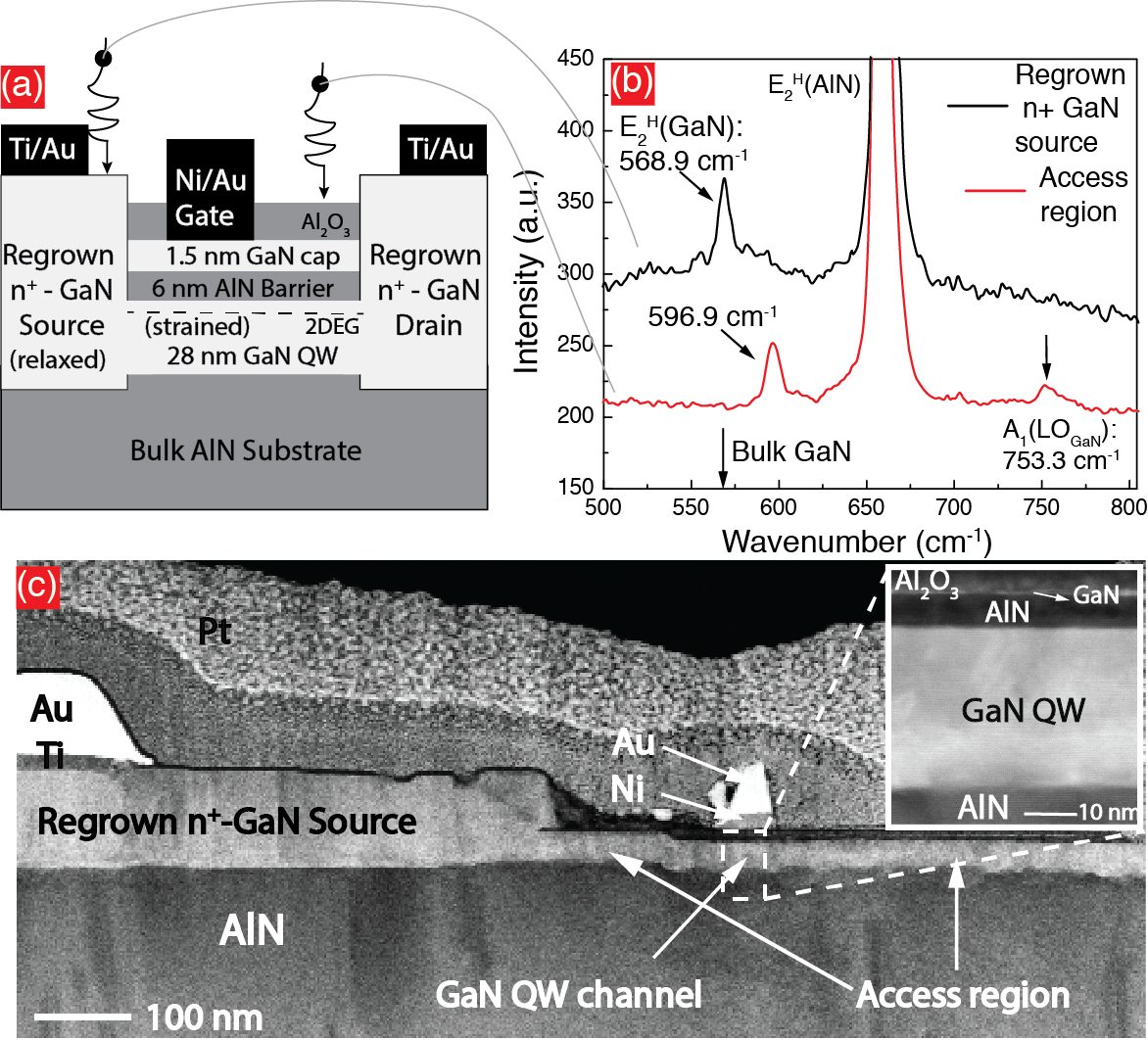}
\caption{\label{fig2} (a) Schematic cross-section of the AlN/GaN/AlN FET consisting a 28-nm-thick GaN QW channel grown on bulk single-crystal AlN substrate. The measured locations by Raman spectroscopy are marked. (b) Raman spectroscopy as phonon probe measured for the access region, and the regrown source region.  The Raman peak of the thick regrown GaN contact region is essentially identical to the bulk GaN peak shown by the arrow showing minimal residual strain.  The thinner GaN QW channel in the access region is however significantly strained as indicated by the large shift in the peak.  (c) $\mathcal{Z}$-contrast cross-section scanning transmission electron microscopy (STEM) image showing the structure of the FET in the dashed region in (a).  The inset image shows the high resolution STEM of the gated GaN QW channel region.}
\end{figure}

An MBE regrowth process was employed to form n+ GaN ohmic contacts to the 2DEG channel using a technique similar to what we have reported earlier\cite{pss11_guo_regrown_hemt, apl12_faria_doping_contacts, edl12_regrown_contacts}.  A $\sim$200-nm-thick SiO$_2$ mask was deposited on the nitride heterostructures using plasma-enhanced chemical vapor deposition (PECVD) and then patterned by reactive ion etching (RIE).  The nitride regrowth regions were etched for $\sim$40 nm using low-damage BCl$_3$/Cl$_2$ inductively coupled plasma reactive ion etching (ICP-RIE).  MBE regrowth of $\sim$100-nm-thick heavily Si-doped ($N_D \sim 10^{20}$/cm$^3$) n+ GaN was performed.  The polycrystalline GaN deposited on the SiO$_2$ mask was lifted off selectively.  Nonalloyed ohmic contacts were formed by E-beam evaporation of 20/100 nm Ti/Au stacks.  Optical lithography was used to pattern and deposit 50/100 nm Ni/Au gates for long channel devices directly on the GaN cap layer as indicated in Fig \ref{fig2}(a), as well as on top of the Ti/Au layers as contacts.  Before fabricating short gate length transistors, a 5 nm Al$_2$O$_3$ layer was deposited by atomic layer deposition (ALD) on the entire sample.  Ni/Au (15/50 nm) gate stacks were defined by electron-beam lithography (EBL) and lift-off for short gate length devices on top of the ALD oxide, as indicated in Fig \ref{fig1}(a).  A Scanning Electron Microscope (SEM) image of a finished device is shown in Fig. \ref{fig1}(c).  The whole processed sample on the bulk AlN wafer is also shown in Fig. \ref{fig1}(b).  Unlike Sapphire, the bulk AlN crystal is not completely transparent due to the presence of point defects.

The processed HEMT structure allowed the mapping of strain in the GaN quantum well channel using the optical marker technique \cite{apl15_meng_raman}. Fig \ref{fig2}(a) indicates the locations where the Raman spectra shown in Fig \ref{fig2}(b) were measured. Fig \ref{fig2}(c) shows the source-side cross section of the processed device with a zoomed in image of the strained GaN quantum well under the gate.  Because the regrown GaN regions are significantly thicker than the quantum well channel region, the difference in strain should be detectable in Raman spectroscopy.  The measured $E_{2}^{H}$ Raman peaks in the two regions shown in Fig \ref{fig2}(b), reveals the difference in the biaxial strain between these regions. The GaN quantum well region shows the expected peak at $E_{2}^{H} (QW) = 596.9$ cm$^{-1}$, indicating that the quantum well GaN is compressively strained.  The peak frequency and the FWHM of the Raman peaks are similar to the epitaxial heterostructure before the FET device fabrication, indicating the FET device processing preserved the crystal quality in the thin GaN quantum well channel.  The thicker n+ GaN regrown region shows a peak at $E_{2}^{H}$ (n+ GaN) = $568.9$ cm$^{-1}$, which is close to the bulk GaN peak of $E_{2}^{H}$ (Bulk GaN) = $567.4$ cm$^{-1}$.  The small residual compressive strain ($< 0.2$\%) and high relaxation in the n+ regrown GaN regions is expected, because the thickness of ~100 nm is large enough to relax the lattice mismatch almost completely in the heterostructure.


After the fabrication of the field-effect transistor, a 2DEG density of $n_s \sim 3.4 \times 10^{13}$/cm$^2$, an electron mobility of $\mu \sim 180$ cm$^2$/V$\cdot$s, and a sheet resistance of $R_{sh} \sim 1020$ $\Omega/\square$ were extracted by Hall effect measurement at room temperature.  The slight degradation from the as-grown sample may be due to the modification of surface states by the ALD Al$_2$O$_3$ layer.  A contact resistance of 0.13 $\Omega \cdot$mm and a sheet resistance of 1100 $\Omega/\square$  were measured by an independent transmission-line method (TLM); the TLM sheet resistance was in good agreement with the the value obtained from the Hall effect measurement.

\begin{figure}[t]
\includegraphics[scale=0.2]{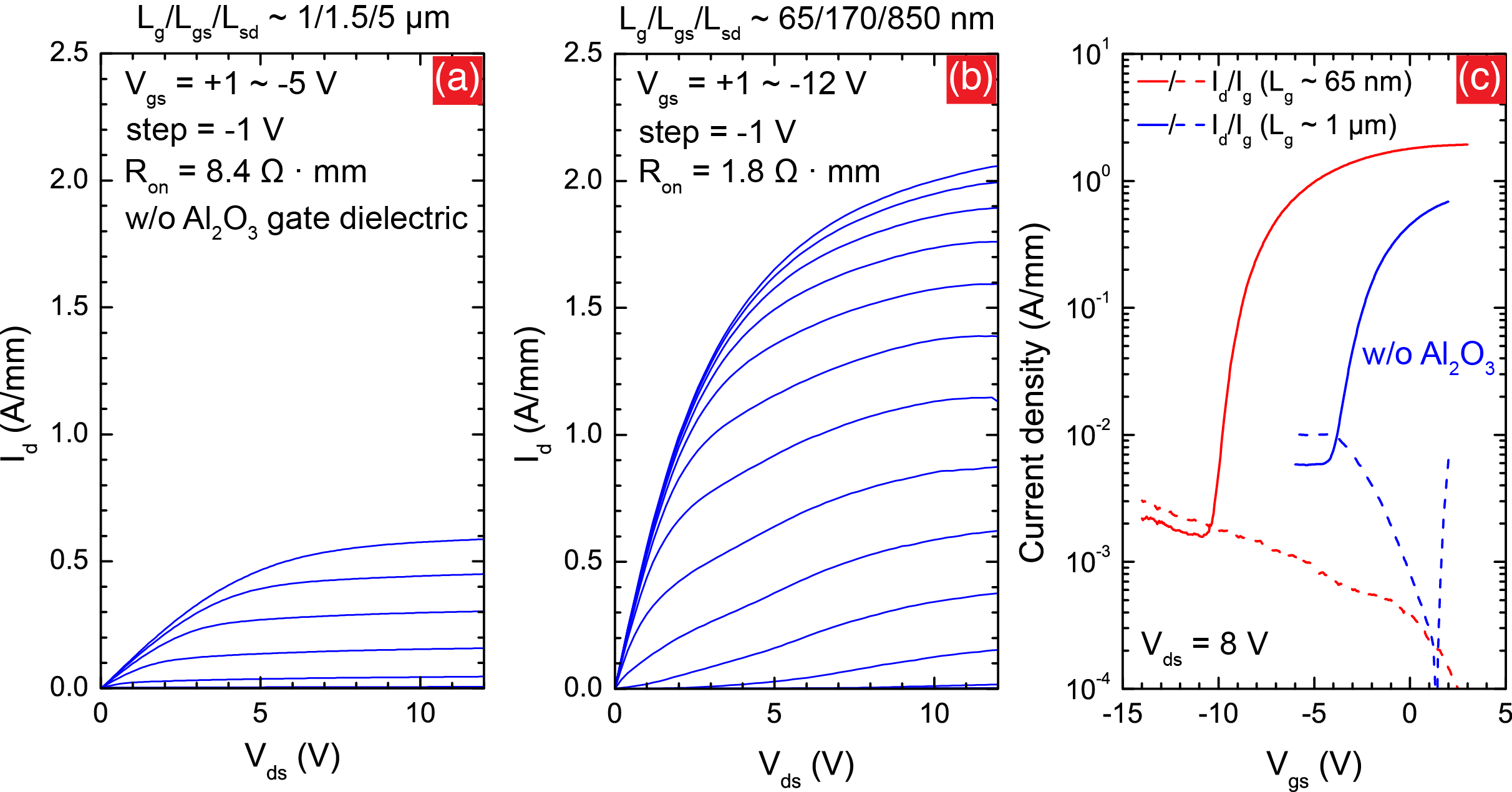}
\caption{\label{fig3} DC common-source family of $I - V$Õs for (a) $L_g$ = 1 $\mu$m and (b) $L_{g}$ = 65 nm AlN/GaN/AlN FETs.  (c) Transfer characteristics of both FETs in semi-log scale.}
\end{figure}

The measured DC common-source output characteristic current-voltage $I - V$'s of (a) a long channel FET and (b) a short channel AlN/GaN/AlN FET are shown in Fig. \ref{fig3}.  The device dimensions are: (a) $W_{g}/L_{g} = 50/1 \mu$m, $L_{gs}/L_{sd} = 1.5/5 \mu$m, and (b) $W_{g}/L_{g} = 2 \times 50 \mu$m/65 nm, $L_{gs}/L_{sd} = 170/850$ nm.  At $V_{gs}$ = +1 V and $V_{ds} = 12$ V, the maximum drain current density for the short channel FETs reaches $I_{d}^{max} \sim 2.0$ A/mm, which is $\sim$3 times higher than the 0.6 A/mm measured for the long channel FETs.  A 80-nm-long gate device exhibited drain current density of 2.8 A/mm at $V_{gs}$ = +3 V and $V_{ds} = 12$ V.  The reduction of the device dimensions thus dramatically boosts the current drive in spite of the high sheet resistance.  

The current drive exceeding 2 A/mm is comparable to state-of-the-art GaN HEMTs on various other substrate platforms \cite{iedm12_shinohara_hemt, edl12_yue_ganhemt_370GHz, edl12_palacios_ganhemt_300GHz, jjap13_yuanzheng_400GHz, drc12_denninghof_ganhemt_400GHz}.  It is a significant improvement from the $\sim$1.4 A/mm of the similar AlN/GaN/AlN FETs that were demonstrated earlier on an AlN-on-sapphire platform \cite{apl14_guowang_gan_pHEMT}.  The on-resistance $R_{on}$ was extracted to be $\sim$ 8.4 $\Omega \cdot $mm and $\sim$1.8 $\Omega \cdot $mm at $V_{gs}$ = +1 V for long and short channel FETs, respectively.  For the 65-nm-long gate devices, a high output conductance was observed for the 28-nm-thick GaN QW channel due to short-channel effects.  With thinner GaN quantum wells and thinner gate barrier stacks, such short channel effects can be suppressed in the future by the presence of the AlN back barrier.  

The on/off transfer characteristics of the long and short channel FETs are shown in Fig. \ref{fig3}(c) in a semi-log scale.  The gate leakage current in the long-channel FET indicates that the MBE growth is not optimal yet, and defects have been generated in the epitaxial process.  The ALD Al$_2$O$_3$ gate dielectric cuts down the leakage, improving the on/off ratio to three orders of magnitude for the short channel devices.  

\begin{figure}[t]
\includegraphics[scale=0.2]{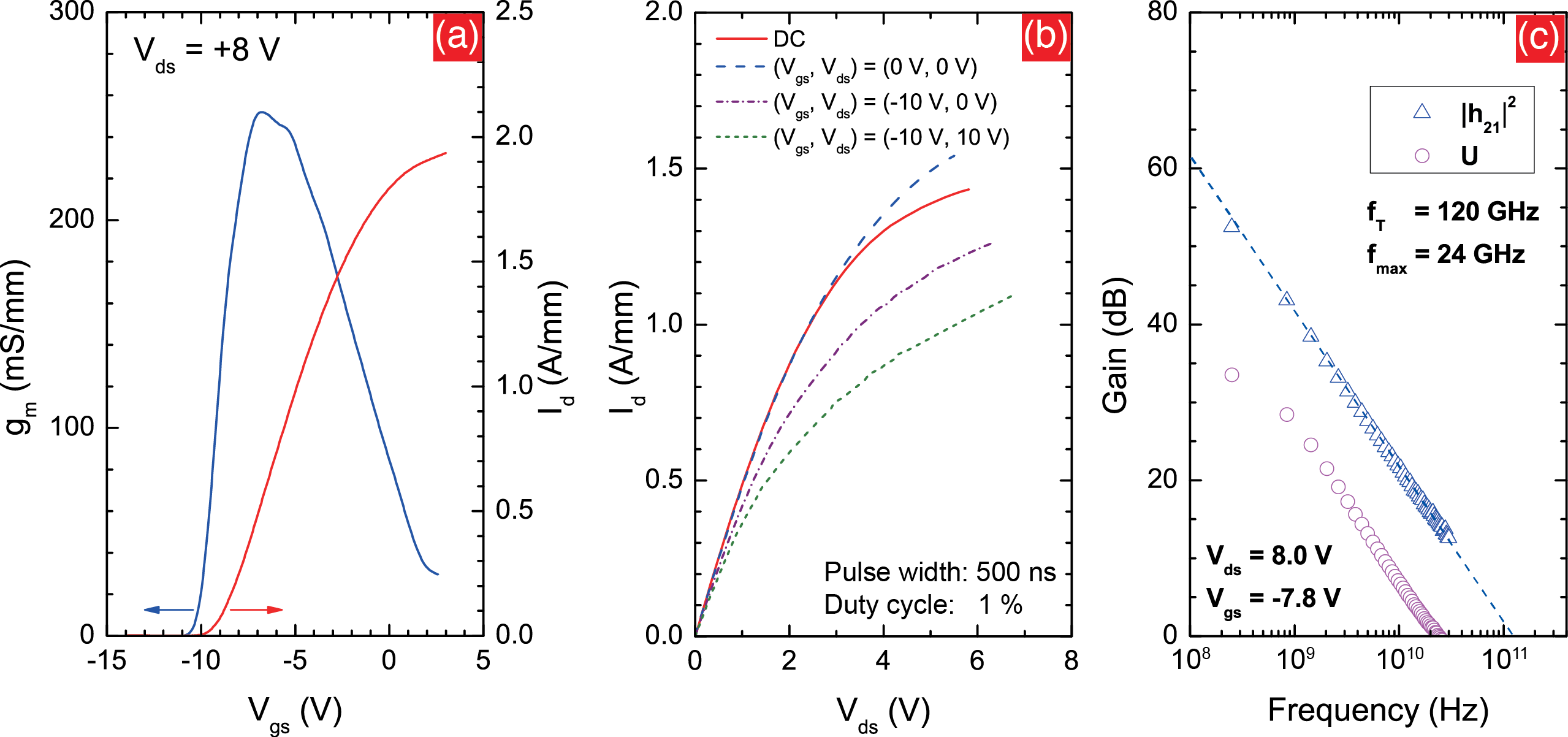}
\caption{\label{fig4} (a) Transfer characteristics of the FETs with 65-nm-long gates in linear scale.  (b) Pulsed IÐV measurements with a 500-ns pulse width and a 0.5-ms period.  (c) Small-signal RF characteristics of the device showing current gain and unilateral gain  $f_T$ / $f_{max}$  = 120/24 GHz.}
\end{figure}

The linear plot of the transistor gain (transconductance) and the transfer characteristics of the 65-nm-long gate device are shown in Fig. \ref{fig4}(a) as a  function of the gate voltage.  The peak extrinsic transconductance reached is $g_{m}^{max} \sim 250$ mS/mm at $V_{gs} = -6.8$ V and $V_{ds}$ = 8 V, corresponding to an intrinsic transconductance of 270 mS/mm for a source access resistance $R_{s} \sim$ 0.3 $\Omega \cdot$mm.  The threshold voltage $V_{th}$ = -9.1 V at $V_{ds}$ = 8 V was obtained by a linear extrapolation of the drain current at the peak transconductance.  The negative threshold voltage is close to what is expected for the high 2DEG density ($3.4 \times10^{13}$/cm$^{2}$) and the gate capacitance.  

The device performance under large-signal drive was investigated by pulsed $I - V$ measurements as shown in Fig. \ref{fig4}(b). Using 500-ns pulse width and a 0.5-ms period, the drain current density $I_{d}$ at $V_{gs}$ = 0 V cold pulsed from ($V_{gs}, V_{ds}$) = (0 V, 0 V) is observed to be higher than what is measured without pulsing at DC.  The drain current drive pulsed from (-10 V, 0 V) and (-10 V, 10 V) showed a $\sim$18\% gate lag and a $\sim$16\% drain lag on the unpassivated devices.  Current saturation is not observed in the pulsed $I - V$ measurements due to short channel effects.  To improve the large signal performance and enhance environmental robustness, passivation of surface states is necessary in future devices\cite{edl11_ronghua_ganhemt_210GHz, edl00_brucegreen_ganhemt_passivation}.   

On-wafer device RF measurements were taken using an HP 8510C vector network analyzer in the frequency range from 0.25 to 30 GHz.    Fig. \ref{fig4}(c) shows the current gain $|h_{21}|^{2}$ and the unilateral gain $\mathcal{U}$ as a function of the frequency for devices biased at peak $f_{T}$ conditions ($V_{gs}$ = -7.8 V, $V_{ds}$ = 8.0 V).  An unity current-gain cutoff frequency $f_{T} = 120$ GHz was extracted by the extrapolation of $|h_{21}|^{2}$ following the 20-dB/decade slope.  Improving the mobility from the current value of $\sim$180 cm$^{2}$/V$\cdot$s will reduce the channel charging delay and enable a higher $f_{T}$ \cite{Moll88}.  An $f_{T} \cdot L_{g}$ product of 7.8 GHz$\cdot \mu$m was obtained, with for a gate-length-to-barrier-thickness aspect ratio of 5.2.  For similar AlN/GaN/AlN FETs without a gate dielectric on AlN template on sapphire \cite{apl14_guowang_gan_pHEMT}, the $f_{T} \cdot L_{g}$ product is 10.4 GHz$\cdot \mu$m with aspect ratio of 13.3.  Due to resistive rectangular gates, the measured low $f_{max} \sim 24$ GHz can be enhanced with T-gates \cite{edl13_ronghua_hemt_230_300GHz} in future embodiments of the device.


In conclusion, this work reports DC and RF performance of AlN/GaN/AlN quantum well FETs on bulk AlN substrates with regrown ohmic contacts for the first time.  By performing Raman spectroscopy with the AlN/GaN/AlN quantum wells as an optical marker, we demonstrate direct measurement of significant lateral strain variations between the GaN QW channel region and the regrown GaN region in the fabricated FET devices.  With a 2DEG density of $3.4 \times 10^{13}$/cm$^2$, the devices pinched off and a record high drain current exceeding 2 A/mm was achieved.  In spite of the low electron mobility, the 65-nm-long gate devices show an $f_{T} \sim 120 $ GHz.  As the homoepitaxial material quality improves, boosts in the electron mobility to the level in conventional GaN HEMTs (i.e. $\sim$1500 cm$^2$/V$\cdot$s) is possible, which will result in a major boost in the performance of FETs on bulk AlN.  The HEMT performance stands to benefit tremendously from the high thermal conductivity of the AlN substrate, and by improved lateral and vertical scaling afforded by the AlN substrate platform.  By using thicker large bandgap AlN barrier layers and AlGaN channels, FETs on bulk AlN can significantly improve the breakdown characteristics and thermal handling over the existing state-of-the-art, and can present a compelling case for high power applications.  

This work was supported in part by an Office of Naval Research THz MURI project supported by Dr. Paul Maki, and by the Designing Materials to Revolutionize and Engineer our Future (DMREF) Program under Award Number 1534303, funded by the National Science Foundation.  

\bibliography{dj_bib} 

\begin{thebibliography}{19}%
\makeatletter
\providecommand \@ifxundefined [1]{%
 \@ifx{#1\undefined}
}%
\providecommand \@ifnum [1]{%
 \ifnum #1\expandafter \@firstoftwo
 \else \expandafter \@secondoftwo
 \fi
}%
\providecommand \@ifx [1]{%
 \ifx #1\expandafter \@firstoftwo
 \else \expandafter \@secondoftwo
 \fi
}%
\providecommand \natexlab [1]{#1}%
\providecommand \enquote  [1]{``#1''}%
\providecommand \bibnamefont  [1]{#1}%
\providecommand \bibfnamefont [1]{#1}%
\providecommand \citenamefont [1]{#1}%
\providecommand \href@noop [0]{\@secondoftwo}%
\providecommand \href [0]{\begingroup \@sanitize@url \@href}%
\providecommand \@href[1]{\@@startlink{#1}\@@href}%
\providecommand \@@href[1]{\endgroup#1\@@endlink}%
\providecommand \@sanitize@url [0]{\catcode `\\12\catcode `\$12\catcode
  `\&12\catcode `\#12\catcode `\^12\catcode `\_12\catcode `\%12\relax}%
\providecommand \@@startlink[1]{}%
\providecommand \@@endlink[0]{}%
\providecommand \url  [0]{\begingroup\@sanitize@url \@url }%
\providecommand \@url [1]{\endgroup\@href {#1}{\urlprefix }}%
\providecommand \urlprefix  [0]{URL }%
\providecommand \Eprint [0]{\href }%
\providecommand \doibase [0]{http://dx.doi.org/}%
\providecommand \selectlanguage [0]{\@gobble}%
\providecommand \bibinfo  [0]{\@secondoftwo}%
\providecommand \bibfield  [0]{\@secondoftwo}%
\providecommand \translation [1]{[#1]}%
\providecommand \BibitemOpen [0]{}%
\providecommand \bibitemStop [0]{}%
\providecommand \bibitemNoStop [0]{.\EOS\space}%
\providecommand \EOS [0]{\spacefactor3000\relax}%
\providecommand \BibitemShut  [1]{\csname bibitem#1\endcsname}%
\let\auto@bib@innerbib\@empty
\bibitem [{\citenamefont {Shinohara}\ \emph {et~al.}(2012)\citenamefont
  {Shinohara}, \citenamefont {Regan}, \citenamefont {Corrion}, \citenamefont
  {Brown}, \citenamefont {Tang}, \citenamefont {Wong}, \citenamefont {Candia},
  \citenamefont {Schmitz}, \citenamefont {Fung}, \citenamefont {Kim},\ and\
  \citenamefont {Micovic}}]{iedm12_shinohara_hemt}%
  \BibitemOpen
  \bibfield  {author} {\bibinfo {author} {\bibfnamefont {K.}~\bibnamefont
  {Shinohara}}, \bibinfo {author} {\bibfnamefont {D.}~\bibnamefont {Regan}},
  \bibinfo {author} {\bibfnamefont {A.}~\bibnamefont {Corrion}}, \bibinfo
  {author} {\bibfnamefont {D.}~\bibnamefont {Brown}}, \bibinfo {author}
  {\bibfnamefont {Y.}~\bibnamefont {Tang}}, \bibinfo {author} {\bibfnamefont
  {J.}~\bibnamefont {Wong}}, \bibinfo {author} {\bibfnamefont {G.}~\bibnamefont
  {Candia}}, \bibinfo {author} {\bibfnamefont {A.}~\bibnamefont {Schmitz}},
  \bibinfo {author} {\bibfnamefont {H.}~\bibnamefont {Fung}}, \bibinfo {author}
  {\bibfnamefont {S.}~\bibnamefont {Kim}}, \ and\ \bibinfo {author}
  {\bibfnamefont {M.}~\bibnamefont {Micovic}},\ }\bibfield  {title} {\enquote
  {\bibinfo {title} {{Self-aligned-gate GaN-HEMTs with heavily-doped n+ GaN
  ohmic contacts to 2DEG}},}\ }\href@noop {} {\bibfield  {journal} {\bibinfo
  {journal} {Proc. IEEE IEDM}\ ,\ \bibinfo {pages} {617}} (\bibinfo {year}
  {2012})}\BibitemShut {NoStop}%
\bibitem [{\citenamefont {Yue}\ \emph {et~al.}(2012)\citenamefont {Yue},
  \citenamefont {Hu}, \citenamefont {Guo}, \citenamefont {Sensale-Rodriguez},
  \citenamefont {Li}, \citenamefont {Wang}, \citenamefont {Faria},
  \citenamefont {Fang}, \citenamefont {Song}, \citenamefont {Guo},
  \citenamefont {Kosel}, \citenamefont {Snider}, \citenamefont {Fay},
  \citenamefont {Jena},\ and\ \citenamefont {Xing}}]{edl12_yue_ganhemt_370GHz}%
  \BibitemOpen
  \bibfield  {author} {\bibinfo {author} {\bibfnamefont {Y.}~\bibnamefont
  {Yue}}, \bibinfo {author} {\bibfnamefont {Z.}~\bibnamefont {Hu}}, \bibinfo
  {author} {\bibfnamefont {J.}~\bibnamefont {Guo}}, \bibinfo {author}
  {\bibfnamefont {B.}~\bibnamefont {Sensale-Rodriguez}}, \bibinfo {author}
  {\bibfnamefont {G.}~\bibnamefont {Li}}, \bibinfo {author} {\bibfnamefont
  {R.}~\bibnamefont {Wang}}, \bibinfo {author} {\bibfnamefont {F.}~\bibnamefont
  {Faria}}, \bibinfo {author} {\bibfnamefont {T.}~\bibnamefont {Fang}},
  \bibinfo {author} {\bibfnamefont {B.}~\bibnamefont {Song}}, \bibinfo {author}
  {\bibfnamefont {S.}~\bibnamefont {Guo}}, \bibinfo {author} {\bibfnamefont
  {T.}~\bibnamefont {Kosel}}, \bibinfo {author} {\bibfnamefont
  {G.}~\bibnamefont {Snider}}, \bibinfo {author} {\bibfnamefont
  {P.}~\bibnamefont {Fay}}, \bibinfo {author} {\bibfnamefont {D.}~\bibnamefont
  {Jena}}, \ and\ \bibinfo {author} {\bibfnamefont {H.}~\bibnamefont {Xing}},\
  }\bibfield  {title} {\enquote {\bibinfo {title} {{InAlN/AlN/GaN HEMTs with
  regrown ohmics and f$_T$ of 370 GHz}},}\ }\href@noop {} {\bibfield  {journal}
  {\bibinfo  {journal} {IEEE Electron Dev. Lett.}\ }\textbf {\bibinfo {volume}
  {33}},\ \bibinfo {pages} {988} (\bibinfo {year} {2012})}\BibitemShut
  {NoStop}%
\bibitem [{\citenamefont {Lee}\ \emph {et~al.}(2012)\citenamefont {Lee},
  \citenamefont {Gao}, \citenamefont {Guo}, \citenamefont {Kopp}, \citenamefont
  {Fay},\ and\ \citenamefont {Palacios}}]{edl12_palacios_ganhemt_300GHz}%
  \BibitemOpen
  \bibfield  {author} {\bibinfo {author} {\bibfnamefont {D.~S.}\ \bibnamefont
  {Lee}}, \bibinfo {author} {\bibfnamefont {X.}~\bibnamefont {Gao}}, \bibinfo
  {author} {\bibfnamefont {S.}~\bibnamefont {Guo}}, \bibinfo {author}
  {\bibfnamefont {D.}~\bibnamefont {Kopp}}, \bibinfo {author} {\bibfnamefont
  {P.}~\bibnamefont {Fay}}, \ and\ \bibinfo {author} {\bibfnamefont
  {T.}~\bibnamefont {Palacios}},\ }\bibfield  {title} {\enquote {\bibinfo
  {title} {{300-GHz InAlN/GaN HEMTs with InGaN back barrier}},}\ }\href@noop {}
  {\bibfield  {journal} {\bibinfo  {journal} {IEEE Electron Dev. Lett.}\
  }\textbf {\bibinfo {volume} {32}},\ \bibinfo {pages} {1525} (\bibinfo {year}
  {2012})}\BibitemShut {NoStop}%
\bibitem [{\citenamefont {Yue}\ \emph {et~al.}(2013)\citenamefont {Yue},
  \citenamefont {Hu}, \citenamefont {Guo}, \citenamefont {Sensale-Rodriguez},
  \citenamefont {Li}, \citenamefont {Wang}, \citenamefont {Faria},
  \citenamefont {Song}, \citenamefont {Gao}, \citenamefont {Guo}, \citenamefont
  {Kosel}, \citenamefont {Snider}, \citenamefont {Fay}, \citenamefont {Jena},\
  and\ \citenamefont {Xing}}]{jjap13_yuanzheng_400GHz}%
  \BibitemOpen
  \bibfield  {author} {\bibinfo {author} {\bibfnamefont {Y.}~\bibnamefont
  {Yue}}, \bibinfo {author} {\bibfnamefont {Z.}~\bibnamefont {Hu}}, \bibinfo
  {author} {\bibfnamefont {J.}~\bibnamefont {Guo}}, \bibinfo {author}
  {\bibfnamefont {B.}~\bibnamefont {Sensale-Rodriguez}}, \bibinfo {author}
  {\bibfnamefont {G.}~\bibnamefont {Li}}, \bibinfo {author} {\bibfnamefont
  {R.}~\bibnamefont {Wang}}, \bibinfo {author} {\bibfnamefont {F.}~\bibnamefont
  {Faria}}, \bibinfo {author} {\bibfnamefont {B.}~\bibnamefont {Song}},
  \bibinfo {author} {\bibfnamefont {X.}~\bibnamefont {Gao}}, \bibinfo {author}
  {\bibfnamefont {S.}~\bibnamefont {Guo}}, \bibinfo {author} {\bibfnamefont
  {T.}~\bibnamefont {Kosel}}, \bibinfo {author} {\bibfnamefont
  {G.}~\bibnamefont {Snider}}, \bibinfo {author} {\bibfnamefont
  {P.}~\bibnamefont {Fay}}, \bibinfo {author} {\bibfnamefont {D.}~\bibnamefont
  {Jena}}, \ and\ \bibinfo {author} {\bibfnamefont {H.~G.}\ \bibnamefont
  {Xing}},\ }\bibfield  {title} {\enquote {\bibinfo {title} {{Ultrascaled
  InAlN/GaN High Electron Mobility Transistors with Cutoff Frequency of 400
  GHz}},}\ }\href@noop {} {\bibfield  {journal} {\bibinfo  {journal} {Jap. J.
  Appl. Phys.}\ }\textbf {\bibinfo {volume} {52}},\ \bibinfo {pages} {08JN14}
  (\bibinfo {year} {2013})}\BibitemShut {NoStop}%
\bibitem [{\citenamefont {Denninghoff}\ \emph {et~al.}(2012)\citenamefont
  {Denninghoff}, \citenamefont {Lu}, \citenamefont {Ahmadi}, \citenamefont
  {Keller},\ and\ \citenamefont {Mishra}}]{drc12_denninghof_ganhemt_400GHz}%
  \BibitemOpen
  \bibfield  {author} {\bibinfo {author} {\bibfnamefont {D.~J.}\ \bibnamefont
  {Denninghoff}}, \bibinfo {author} {\bibfnamefont {J.}~\bibnamefont {Lu}},
  \bibinfo {author} {\bibfnamefont {E.}~\bibnamefont {Ahmadi}}, \bibinfo
  {author} {\bibfnamefont {S.}~\bibnamefont {Keller}}, \ and\ \bibinfo {author}
  {\bibfnamefont {U.~K.}\ \bibnamefont {Mishra}},\ }\bibfield  {title}
  {\enquote {\bibinfo {title} {{N-polar GaN/InAlN MIS-HEMT with 400-GHz
  f$_{max}$}},}\ }\href@noop {} {\bibfield  {journal} {\bibinfo  {journal}
  {IEEE Dev. Research Conf. Tech Digest}\ ,\ \bibinfo {pages} {151}} (\bibinfo
  {year} {2012})}\BibitemShut {NoStop}%
\bibitem [{\citenamefont {Slack}\ \emph {et~al.}(1987)\citenamefont {Slack},
  \citenamefont {Tanzilli}, \citenamefont {Pohl},\ and\ \citenamefont
  {Vandersande}}]{jpcm87_slack_aln_thermal_conductivity_340WmK}%
  \BibitemOpen
  \bibfield  {author} {\bibinfo {author} {\bibfnamefont {G.~A.}\ \bibnamefont
  {Slack}}, \bibinfo {author} {\bibfnamefont {R.~A.}\ \bibnamefont {Tanzilli}},
  \bibinfo {author} {\bibfnamefont {R.~O.}\ \bibnamefont {Pohl}}, \ and\
  \bibinfo {author} {\bibfnamefont {J.~W.}\ \bibnamefont {Vandersande}},\
  }\href@noop {} {\bibfield  {journal} {\bibinfo  {journal} {J. Phys. Chem.
  Solids}\ }\textbf {\bibinfo {volume} {48}},\ \bibinfo {pages} {641} (\bibinfo
  {year} {1987})}\BibitemShut {NoStop}%
\bibitem [{\citenamefont {Levinshtein~M.E.}(2001)}]{shur_gan_properties}%
  \BibitemOpen
  \bibfield  {author} {\bibinfo {author} {\bibfnamefont {S.~M.~E.}\
  \bibnamefont {Levinshtein~M.E.}, \bibfnamefont {Rumyantsev~S.L.}},\
  }\href@noop {} {\emph {\bibinfo {title} {Properties of Advanced Semiconductor
  Materials: GaN, AlN, SiC, BN, SiC, SiGe}}}\ (\bibinfo  {publisher} {John
  Wiley \& Sons},\ \bibinfo {address} {New Jersey},\ \bibinfo {year}
  {2001})\BibitemShut {NoStop}%
\bibitem [{\citenamefont {Hu}\ \emph {et~al.}(2003)\citenamefont {Hu},
  \citenamefont {Deng}, \citenamefont {Pala}, \citenamefont {Gaska},
  \citenamefont {Shur}, \citenamefont {Chen}, \citenamefont {Yang},
  \citenamefont {Simin}, \citenamefont {Khan}, \citenamefont {Rojo},\ and\
  \citenamefont {Schowalter}}]{apl03_gan_hemt_on_aln_crystal_IS}%
  \BibitemOpen
  \bibfield  {author} {\bibinfo {author} {\bibfnamefont {X.}~\bibnamefont
  {Hu}}, \bibinfo {author} {\bibfnamefont {J.}~\bibnamefont {Deng}}, \bibinfo
  {author} {\bibfnamefont {N.}~\bibnamefont {Pala}}, \bibinfo {author}
  {\bibfnamefont {R.}~\bibnamefont {Gaska}}, \bibinfo {author} {\bibfnamefont
  {M.~S.}\ \bibnamefont {Shur}}, \bibinfo {author} {\bibfnamefont {C.~Q.}\
  \bibnamefont {Chen}}, \bibinfo {author} {\bibfnamefont {J.}~\bibnamefont
  {Yang}}, \bibinfo {author} {\bibfnamefont {G.}~\bibnamefont {Simin}},
  \bibinfo {author} {\bibfnamefont {M.~A.}\ \bibnamefont {Khan}}, \bibinfo
  {author} {\bibfnamefont {J.~C.}\ \bibnamefont {Rojo}}, \ and\ \bibinfo
  {author} {\bibfnamefont {L.~J.}\ \bibnamefont {Schowalter}},\ }\bibfield
  {title} {\enquote {\bibinfo {title} {{AlGaN/GaN heterostructure field-effect
  transistors on single-crystal bulk AlN}},}\ }\href@noop {} {\bibfield
  {journal} {\bibinfo  {journal} {Appl. Phys. Lett.}\ }\textbf {\bibinfo
  {volume} {82}},\ \bibinfo {pages} {1299} (\bibinfo {year}
  {2003})}\BibitemShut {NoStop}%
\bibitem [{\citenamefont {Qi}\ \emph {et~al.}(2015)\citenamefont {Qi},
  \citenamefont {Li}, \citenamefont {Protasenko}, \citenamefont {Zhao},
  \citenamefont {Verma}, \citenamefont {Song}, \citenamefont {Ganguly},
  \citenamefont {Zhu}, \citenamefont {Hu}, \citenamefont {Yan}, \citenamefont
  {Mintairov}, \citenamefont {Xing},\ and\ \citenamefont
  {Jena}}]{apl15_meng_raman}%
  \BibitemOpen
  \bibfield  {author} {\bibinfo {author} {\bibfnamefont {M.}~\bibnamefont
  {Qi}}, \bibinfo {author} {\bibfnamefont {G.}~\bibnamefont {Li}}, \bibinfo
  {author} {\bibfnamefont {V.}~\bibnamefont {Protasenko}}, \bibinfo {author}
  {\bibfnamefont {P.}~\bibnamefont {Zhao}}, \bibinfo {author} {\bibfnamefont
  {J.}~\bibnamefont {Verma}}, \bibinfo {author} {\bibfnamefont
  {B.}~\bibnamefont {Song}}, \bibinfo {author} {\bibfnamefont {S.}~\bibnamefont
  {Ganguly}}, \bibinfo {author} {\bibfnamefont {M.}~\bibnamefont {Zhu}},
  \bibinfo {author} {\bibfnamefont {Z.}~\bibnamefont {Hu}}, \bibinfo {author}
  {\bibfnamefont {X.}~\bibnamefont {Yan}}, \bibinfo {author} {\bibfnamefont
  {A.}~\bibnamefont {Mintairov}}, \bibinfo {author} {\bibfnamefont
  {H.}~\bibnamefont {Xing}}, \ and\ \bibinfo {author} {\bibfnamefont
  {D.}~\bibnamefont {Jena}},\ }\bibfield  {title} {\enquote {\bibinfo {title}
  {{Dual optical marker Raman characterization of strained GaN-channels on AlN
  using AlN/GaN/AlN quantum wells and $^{15}$N isotopes}},}\ }\href@noop {}
  {\bibfield  {journal} {\bibinfo  {journal} {Appl. Phys. Lett.}\ }\textbf
  {\bibinfo {volume} {106}},\ \bibinfo {pages} {106} (\bibinfo {year}
  {2015})}\BibitemShut {NoStop}%
\bibitem [{\citenamefont {Li}\ \emph {et~al.}(2014)\citenamefont {Li},
  \citenamefont {Song}, \citenamefont {Ganguly}, \citenamefont {Zhu},
  \citenamefont {Wang}, \citenamefont {Yan}, \citenamefont {Verma},
  \citenamefont {Protasenko}, \citenamefont {Xing},\ and\ \citenamefont
  {Jena}}]{apl14_guowang_gan_pHEMT}%
  \BibitemOpen
  \bibfield  {author} {\bibinfo {author} {\bibfnamefont {G.}~\bibnamefont
  {Li}}, \bibinfo {author} {\bibfnamefont {B.}~\bibnamefont {Song}}, \bibinfo
  {author} {\bibfnamefont {S.}~\bibnamefont {Ganguly}}, \bibinfo {author}
  {\bibfnamefont {M.}~\bibnamefont {Zhu}}, \bibinfo {author} {\bibfnamefont
  {R.}~\bibnamefont {Wang}}, \bibinfo {author} {\bibfnamefont {X.}~\bibnamefont
  {Yan}}, \bibinfo {author} {\bibfnamefont {J.}~\bibnamefont {Verma}}, \bibinfo
  {author} {\bibfnamefont {V.}~\bibnamefont {Protasenko}}, \bibinfo {author}
  {\bibfnamefont {H.}~\bibnamefont {Xing}}, \ and\ \bibinfo {author}
  {\bibfnamefont {D.}~\bibnamefont {Jena}},\ }\bibfield  {title} {\enquote
  {\bibinfo {title} {{Two-dimensional electron gases in strained quantum wells
  for AlN/GaN/AlN double heterostructure field-effect transistors on AlN}},}\
  }\href@noop {} {\bibfield  {journal} {\bibinfo  {journal} {Applied Physics
  Letters}\ }\textbf {\bibinfo {volume} {104}},\ \bibinfo {pages} {1--5}
  (\bibinfo {year} {2014})}\BibitemShut {NoStop}%
\bibitem [{\citenamefont {Li}\ \emph {et~al.}(2012)\citenamefont {Li},
  \citenamefont {Wang}, \citenamefont {Guo}, \citenamefont {Verma},
  \citenamefont {Hu}, \citenamefont {Yue}, \citenamefont {Faria}, \citenamefont
  {Cao}, \citenamefont {Kelly}, \citenamefont {Kosel}, \citenamefont {Xing},\
  and\ \citenamefont {Jena}}]{edl12_guowang_ganQW_nFET}%
  \BibitemOpen
  \bibfield  {author} {\bibinfo {author} {\bibfnamefont {G.}~\bibnamefont
  {Li}}, \bibinfo {author} {\bibfnamefont {R.}~\bibnamefont {Wang}}, \bibinfo
  {author} {\bibfnamefont {J.}~\bibnamefont {Guo}}, \bibinfo {author}
  {\bibfnamefont {J.}~\bibnamefont {Verma}}, \bibinfo {author} {\bibfnamefont
  {Z.}~\bibnamefont {Hu}}, \bibinfo {author} {\bibfnamefont {Y.}~\bibnamefont
  {Yue}}, \bibinfo {author} {\bibfnamefont {F.}~\bibnamefont {Faria}}, \bibinfo
  {author} {\bibfnamefont {Y.}~\bibnamefont {Cao}}, \bibinfo {author}
  {\bibfnamefont {M.}~\bibnamefont {Kelly}}, \bibinfo {author} {\bibfnamefont
  {T.}~\bibnamefont {Kosel}}, \bibinfo {author} {\bibfnamefont
  {H.}~\bibnamefont {Xing}}, \ and\ \bibinfo {author} {\bibfnamefont
  {D.}~\bibnamefont {Jena}},\ }\bibfield  {title} {\enquote {\bibinfo {title}
  {{Ultrathin Body GaN-on-Insulator Quantum Well FETs With Regrown Ohmic
  Contacts}},}\ }\href@noop {} {\bibfield  {journal} {\bibinfo  {journal} {IEEE
  Electron Device Letters}\ }\textbf {\bibinfo {volume} {33}},\ \bibinfo
  {pages} {661--664} (\bibinfo {year} {2012})}\BibitemShut {NoStop}%
\bibitem [{\citenamefont {Jana}\ and\ \citenamefont
  {Jena}(2011)}]{apl11_raj_stark_scattering}%
  \BibitemOpen
  \bibfield  {author} {\bibinfo {author} {\bibfnamefont {R.}~\bibnamefont
  {Jana}}\ and\ \bibinfo {author} {\bibfnamefont {D.}~\bibnamefont {Jena}},\
  }\bibfield  {title} {\enquote {\bibinfo {title} {{Stark-effect scattering in
  rough quantum wells}},}\ }\href@noop {} {\bibfield  {journal} {\bibinfo
  {journal} {Appl. Phys. Lett.}\ }\textbf {\bibinfo {volume} {99}},\ \bibinfo
  {pages} {99} (\bibinfo {year} {2011})}\BibitemShut {NoStop}%
\bibitem [{\citenamefont {Guo}\ \emph {et~al.}(2011)\citenamefont {Guo},
  \citenamefont {Cao}, \citenamefont {Lian}, \citenamefont {Zimmermann},
  \citenamefont {Li}, \citenamefont {Verma}, \citenamefont {Gao}, \citenamefont
  {Guo}, \citenamefont {Saunier}, \citenamefont {Wistey}, \citenamefont
  {Jena},\ and\ \citenamefont {Xing}}]{pss11_guo_regrown_hemt}%
  \BibitemOpen
  \bibfield  {author} {\bibinfo {author} {\bibfnamefont {J.}~\bibnamefont
  {Guo}}, \bibinfo {author} {\bibfnamefont {Y.}~\bibnamefont {Cao}}, \bibinfo
  {author} {\bibfnamefont {C.}~\bibnamefont {Lian}}, \bibinfo {author}
  {\bibfnamefont {T.}~\bibnamefont {Zimmermann}}, \bibinfo {author}
  {\bibfnamefont {G.}~\bibnamefont {Li}}, \bibinfo {author} {\bibfnamefont
  {J.}~\bibnamefont {Verma}}, \bibinfo {author} {\bibfnamefont
  {X.}~\bibnamefont {Gao}}, \bibinfo {author} {\bibfnamefont {S.}~\bibnamefont
  {Guo}}, \bibinfo {author} {\bibfnamefont {P.}~\bibnamefont {Saunier}},
  \bibinfo {author} {\bibfnamefont {M.}~\bibnamefont {Wistey}}, \bibinfo
  {author} {\bibfnamefont {D.}~\bibnamefont {Jena}}, \ and\ \bibinfo {author}
  {\bibfnamefont {H.}~\bibnamefont {Xing}},\ }\bibfield  {title} {\enquote
  {\bibinfo {title} {{Metal-face InAlN/AlN/GaN high electron mobility
  transistors with regrown ohmic contacts by molecular beam epitaxy}},}\
  }\href@noop {} {\bibfield  {journal} {\bibinfo  {journal} {Phys. Stat. Sol.
  (A)}\ }\textbf {\bibinfo {volume} {208}},\ \bibinfo {pages} {1617} (\bibinfo
  {year} {2011})}\BibitemShut {NoStop}%
\bibitem [{\citenamefont {Faria}\ \emph {et~al.}(2012)\citenamefont {Faria},
  \citenamefont {Guo}, \citenamefont {Zhao}, \citenamefont {Li}, \citenamefont
  {Kandaswamy}, \citenamefont {Wistey}, \citenamefont {Xing},\ and\
  \citenamefont {Jena}}]{apl12_faria_doping_contacts}%
  \BibitemOpen
  \bibfield  {author} {\bibinfo {author} {\bibfnamefont {F.}~\bibnamefont
  {Faria}}, \bibinfo {author} {\bibfnamefont {J.}~\bibnamefont {Guo}}, \bibinfo
  {author} {\bibfnamefont {P.}~\bibnamefont {Zhao}}, \bibinfo {author}
  {\bibfnamefont {G.}~\bibnamefont {Li}}, \bibinfo {author} {\bibfnamefont
  {P.}~\bibnamefont {Kandaswamy}}, \bibinfo {author} {\bibfnamefont
  {M.}~\bibnamefont {Wistey}}, \bibinfo {author} {\bibfnamefont
  {H.}~\bibnamefont {Xing}}, \ and\ \bibinfo {author} {\bibfnamefont
  {D.}~\bibnamefont {Jena}},\ }\bibfield  {title} {\enquote {\bibinfo {title}
  {{Ultra-low resistance ohmic contacts to GaN with high Si doping
  concentrations grown by molecular beam epitaxy}},}\ }\href@noop {} {\bibfield
   {journal} {\bibinfo  {journal} {Appl. Phys. Lett.}\ }\textbf {\bibinfo
  {volume} {101}},\ \bibinfo {pages} {032109} (\bibinfo {year}
  {2012})}\BibitemShut {NoStop}%
\bibitem [{\citenamefont {Guo}\ \emph {et~al.}(2012)\citenamefont {Guo},
  \citenamefont {Li}, \citenamefont {Faria}, \citenamefont {Cao}, \citenamefont
  {Wang}, \citenamefont {Verma}, \citenamefont {Gao}, \citenamefont {Guo},
  \citenamefont {Beam}, \citenamefont {Ketterson}, \citenamefont {Schuette},
  \citenamefont {Saunier}, \citenamefont {Wistey}, \citenamefont {Jena},\ and\
  \citenamefont {Xing}}]{edl12_regrown_contacts}%
  \BibitemOpen
  \bibfield  {author} {\bibinfo {author} {\bibfnamefont {J.}~\bibnamefont
  {Guo}}, \bibinfo {author} {\bibfnamefont {G.}~\bibnamefont {Li}}, \bibinfo
  {author} {\bibfnamefont {F.}~\bibnamefont {Faria}}, \bibinfo {author}
  {\bibfnamefont {Y.}~\bibnamefont {Cao}}, \bibinfo {author} {\bibfnamefont
  {R.}~\bibnamefont {Wang}}, \bibinfo {author} {\bibfnamefont {J.}~\bibnamefont
  {Verma}}, \bibinfo {author} {\bibfnamefont {X.}~\bibnamefont {Gao}}, \bibinfo
  {author} {\bibfnamefont {S.}~\bibnamefont {Guo}}, \bibinfo {author}
  {\bibfnamefont {E.}~\bibnamefont {Beam}}, \bibinfo {author} {\bibfnamefont
  {A.}~\bibnamefont {Ketterson}}, \bibinfo {author} {\bibfnamefont
  {M.}~\bibnamefont {Schuette}}, \bibinfo {author} {\bibfnamefont
  {P.}~\bibnamefont {Saunier}}, \bibinfo {author} {\bibfnamefont
  {M.}~\bibnamefont {Wistey}}, \bibinfo {author} {\bibfnamefont
  {D.}~\bibnamefont {Jena}}, \ and\ \bibinfo {author} {\bibfnamefont
  {H.}~\bibnamefont {Xing}},\ }\bibfield  {title} {\enquote {\bibinfo {title}
  {{MBE-Regrown Ohmics in InAlN HEMTs With a Regrowth Interface Resistance of
  0.05 Ohm$\cdot$mm}},}\ }\href@noop {} {\bibfield  {journal} {\bibinfo
  {journal} {IEEE Electron Dev. Lett.}\ }\textbf {\bibinfo {volume} {33}},\
  \bibinfo {pages} {525} (\bibinfo {year} {2012})}\BibitemShut {NoStop}%
\bibitem [{\citenamefont {Wang}\ \emph {et~al.}(2011)\citenamefont {Wang},
  \citenamefont {Li}, \citenamefont {Laboutin}, \citenamefont {Cao},
  \citenamefont {Johnson}, \citenamefont {Snider}, \citenamefont {Fay},
  \citenamefont {Jena},\ and\ \citenamefont
  {Xing}}]{edl11_ronghua_ganhemt_210GHz}%
  \BibitemOpen
  \bibfield  {author} {\bibinfo {author} {\bibfnamefont {R.}~\bibnamefont
  {Wang}}, \bibinfo {author} {\bibfnamefont {G.}~\bibnamefont {Li}}, \bibinfo
  {author} {\bibfnamefont {O.}~\bibnamefont {Laboutin}}, \bibinfo {author}
  {\bibfnamefont {Y.}~\bibnamefont {Cao}}, \bibinfo {author} {\bibfnamefont
  {W.}~\bibnamefont {Johnson}}, \bibinfo {author} {\bibfnamefont
  {G.}~\bibnamefont {Snider}}, \bibinfo {author} {\bibfnamefont
  {P.}~\bibnamefont {Fay}}, \bibinfo {author} {\bibfnamefont {D.}~\bibnamefont
  {Jena}}, \ and\ \bibinfo {author} {\bibfnamefont {H.}~\bibnamefont {Xing}},\
  }\bibfield  {title} {\enquote {\bibinfo {title} {{210-GHz InAlN/GaN HEMTs
  with dielectric-free passivation}},}\ }\href@noop {} {\bibfield  {journal}
  {\bibinfo  {journal} {IEEE Electron Dev. Lett.}\ }\textbf {\bibinfo {volume}
  {32}},\ \bibinfo {pages} {892} (\bibinfo {year} {2011})}\BibitemShut
  {NoStop}%
\bibitem [{\citenamefont {Green}\ \emph {et~al.}(2000)\citenamefont {Green},
  \citenamefont {Chu}, \citenamefont {Chumbes}, \citenamefont {Smart},
  \citenamefont {Shealy},\ and\ \citenamefont
  {Eastman}}]{edl00_brucegreen_ganhemt_passivation}%
  \BibitemOpen
  \bibfield  {author} {\bibinfo {author} {\bibfnamefont {B.~M.}\ \bibnamefont
  {Green}}, \bibinfo {author} {\bibfnamefont {K.~K.}\ \bibnamefont {Chu}},
  \bibinfo {author} {\bibfnamefont {E.~M.}\ \bibnamefont {Chumbes}}, \bibinfo
  {author} {\bibfnamefont {J.~A.}\ \bibnamefont {Smart}}, \bibinfo {author}
  {\bibfnamefont {J.~R.}\ \bibnamefont {Shealy}}, \ and\ \bibinfo {author}
  {\bibfnamefont {L.~F.}\ \bibnamefont {Eastman}},\ }\bibfield  {title}
  {\enquote {\bibinfo {title} {{The effect of surface passivation on the
  microwave characteristics of AlGaN/GaN HFETs}},}\ }\href@noop {} {\bibfield
  {journal} {\bibinfo  {journal} {IEEE Electron Dev. Lett.}\ }\textbf {\bibinfo
  {volume} {21}},\ \bibinfo {pages} {268} (\bibinfo {year} {2000})}\BibitemShut
  {NoStop}%
\bibitem [{\citenamefont {Moll}, \citenamefont {Hueschen},\ and\ \citenamefont
  {Fischer-Colbrie}(1988)}]{Moll88}%
  \BibitemOpen
  \bibfield  {author} {\bibinfo {author} {\bibfnamefont {N.}~\bibnamefont
  {Moll}}, \bibinfo {author} {\bibfnamefont {M.~R.}\ \bibnamefont {Hueschen}},
  \ and\ \bibinfo {author} {\bibfnamefont {A.}~\bibnamefont
  {Fischer-Colbrie}},\ }\bibfield  {title} {\enquote {\bibinfo {title}
  {{Pulse-Doped AlGaAs/InGaAs Pseudomorphic MODFETs}},}\ }\href@noop {}
  {\bibfield  {journal} {\bibinfo  {journal} {IEEE Trans. on Electron. Dev.}\
  }\textbf {\bibinfo {volume} {35}},\ \bibinfo {pages} {879} (\bibinfo {year}
  {1988})}\BibitemShut {NoStop}%
\bibitem [{\citenamefont {Wang}\ \emph {et~al.}(2013)\citenamefont {Wang},
  \citenamefont {Li}, \citenamefont {Karbasian}, \citenamefont {Guo},
  \citenamefont {Song}, \citenamefont {Yue}, \citenamefont {Hu}, \citenamefont
  {Laboutin}, \citenamefont {Cao}, \citenamefont {Johnson}, \citenamefont
  {Snider}, \citenamefont {Fay}, \citenamefont {Jena},\ and\ \citenamefont
  {Xing}}]{edl13_ronghua_hemt_230_300GHz}%
  \BibitemOpen
  \bibfield  {author} {\bibinfo {author} {\bibfnamefont {R.}~\bibnamefont
  {Wang}}, \bibinfo {author} {\bibfnamefont {G.}~\bibnamefont {Li}}, \bibinfo
  {author} {\bibfnamefont {G.}~\bibnamefont {Karbasian}}, \bibinfo {author}
  {\bibfnamefont {J.}~\bibnamefont {Guo}}, \bibinfo {author} {\bibfnamefont
  {B.}~\bibnamefont {Song}}, \bibinfo {author} {\bibfnamefont {Y.}~\bibnamefont
  {Yue}}, \bibinfo {author} {\bibfnamefont {Z.}~\bibnamefont {Hu}}, \bibinfo
  {author} {\bibfnamefont {O.}~\bibnamefont {Laboutin}}, \bibinfo {author}
  {\bibfnamefont {Y.}~\bibnamefont {Cao}}, \bibinfo {author} {\bibfnamefont
  {W.}~\bibnamefont {Johnson}}, \bibinfo {author} {\bibfnamefont
  {G.}~\bibnamefont {Snider}}, \bibinfo {author} {\bibfnamefont
  {P.}~\bibnamefont {Fay}}, \bibinfo {author} {\bibfnamefont {D.}~\bibnamefont
  {Jena}}, \ and\ \bibinfo {author} {\bibfnamefont {H.~G.}\ \bibnamefont
  {Xing}},\ }\bibfield  {title} {\enquote {\bibinfo {title} {{Quaternary
  Barrier InAlGaN HEMTs With Ê$f_{T}/f_{max}$Ê of 230/300 GHz}},}\ }\href@noop
  {} {\bibfield  {journal} {\bibinfo  {journal} {IEEE Electron Device Letters}\
  }\textbf {\bibinfo {volume} {34}},\ \bibinfo {pages} {378--380} (\bibinfo
  {year} {2013})}\BibitemShut {NoStop}%
\end{thebibliography}%

\end{document}